
\input amssym.def
\input amssym.tex

\magnification=1200
\baselineskip=14pt
\tolerance=1000\hfuzz=1pt


\font\bigfont=cmr10 scaled\magstep3

\def\section#1#2{\vskip32pt plus4pt \goodbreak \noindent{\bf#1. #2}
        \xdef\currentsec{#1} \global\eqnum=0 \global\thmnum=0}

\def\leftheader{}
\def\rightheader{}
\def\header{\headline={\tenrm\ifnum\pageno>1
                        \ifodd\pageno\rightheader
                         \else\noindent\leftheader
                          \fi\else\hfil\fi}}
\newcount\thmnum
\global\thmnum=0
\def\prop#1#2{\global\advance\thmnum by 1
        \xdef#1{Proposition \currentsec.\the\thmnum}
        \bigbreak\noindent{\bf Proposition \currentsec.\the\thmnum.}
        {\it#2} }
\def\define#1#2{\global\advance\thmnum by 1
        \xdef#1{Definition \currentsec.\the\thmnum}
        \bigbreak\noindent{\bf Definition \currentsec.\the\thmnum.}
        {\it#2} }
\def\lemma#1#2{\global\advance\thmnum by 1
        \xdef#1{Lemma \currentsec.\the\thmnum}
        \bigbreak\noindent{\bf Lemma \currentsec.\the\thmnum.}
        {\it#2}}
\def\thm#1#2{\global\advance\thmnum by 1
        \xdef#1{Theorem \currentsec.\the\thmnum}
        \bigbreak\noindent{\bf Theorem \currentsec.\the\thmnum.}
        {\it#2} }
\def\cor#1#2{\global\advance\thmnum by 1
        \xdef#1{Corollary \currentsec.\the\thmnum}
        \bigbreak\noindent{\bf Corollary \currentsec.\the\thmnum.}
        {\it#2} }
\def\conj#1#2{\global\advance\thmnum by 1
        \xdef#1{Conjecture \currentsec.\the\thmnum}
        \bigbreak\noindent{\bf Conjecture \currentsec.\the\thmnum.}
        {\it#2} }

\def\proof{\medskip\noindent{\it Proof. }}

\newcount\eqnum
\global\eqnum=0
\def\num{\global\advance\eqnum by 1
        \eqno({\rm\currentsec}.\the\eqnum)}
\def\eqalignnum{\global\advance\eqnum by 1
        ({\rm\currentsec}.\the\eqnum)}
\def\ref#1{\num  \xdef#1{(\currentsec.\the\eqnum)}}
\def\eqalignref#1{\eqalignnum  \xdef#1{(\currentsec.\the\eqnum)}}

\def\title#1{\centerline{\bf\bigfont#1}}

\newcount\subnum
\def\Alph#1{\ifcase#1\or A\or B\or C\or D\or E\or F\or G\or H\fi}
\def\subsec{\global\advance\subnum by 1
        \vskip12pt plus4pt \goodbreak \noindent
        {\bf \currentsec.\Alph\subnum.}  }
\def\newsubsec{\global\subnum=1 \vskip6pt\noindent
        {\bf \currentsec.\Alph\subnum.}  }
\def\today{\ifcase\month\or January\or February\or March\or
        April\or May\or June\or July\or August\or September\or
        October\or November\or December\fi\space\number\day,
        \number\year}

\def\ol{\overline}

\def\tr{\mathop{\rm tr}\nolimits}

\def\dint{\int^\oplus}
\def\bC{{\Bbb C}}

\def\bR{{\Bbb R}}
\def\bT{{\Bbb T}}
\def\bZ{{\Bbb Z}}

\def\cH{{\cal H}}

\def\fA{{\frak A}}

\chardef\o="1C
\def\frac#1#2{{#1\over #2}}

\header
\def\leftheader{\centerline{S. KLIMEK, A. LE\'SNIEWSKI, N. MAITRA,
                            and R. RUBIN}}
\def\rightheader{\centerline{QUANTIZED TORAL AUTOMORPHISMS}}
\def\intsec{I}
\def\torautsec{II}
\def\catsec{III}
\def\kroneckersec{IV}
\def\catstructsec{V}
\def\kronstructsec{VI}

\def\ct{C(\bT^2)}
\def\alg{\fA_\hbar}
\def\hil{\cH^2(\bC,d\mu_\hbar)}
\def\im{{\rm Im}}
\def\re{{\rm Re}}

{\baselineskip=12pt
\nopagenumbers
\line{\hfill \bf HUTMP 95/444}
\line{\hfill December 4, 1995}
\vfill
\title{Ergodic properties}
\vskip 1cm
\title{of quantized toral automorphisms}
\vskip 1in
\centerline{
{\bf S\l awomir Klimek,}$^*$\footnote{$^1$}
{Supported in part by the National Science Foundation under grant
DMS--9500463}
{\bf Andrzej Le\'sniewski,}$^{**}$\footnote{$^2$}
{Supported in part by the National Science Foundation under grant
DMS--9424344 and by the Department of Energy under grant
DE--FG02--88ER25065}
{\bf Neepa Maitra,}$^{**}$\footnote{$^3$}
{Supported in part by the National Science Foundation under grant
CHE-9321260}
and {\bf Ron Rubin}$^{**}$\footnote{$^4$}
{Supported in part by a National Science Foundation Graduate
Research Fellowship}}
\vskip 12pt
\centerline{$^*$Department of Mathematics}
\centerline{IUPUI}
\centerline{Indianapolis, IN 46205, USA}
\vskip 12pt
\centerline{$^{**}$Lyman Laboratory of Physics}
\centerline{Harvard University}
\centerline{Cambridge, MA 02138, USA}
\vskip 1in\noindent
{\centerline{\bf Abstract}}

\noindent
We study the ergodic properties for a class of quantized toral
automorphisms, namely the cat and Kronecker maps. The present
work uses and extends the results of [KL]. We show that
quantized cat maps are strongly mixing, while Kronecker maps
are ergodic and non-mixing. We also study the structure of
these quantum maps and show that they are effected by unitary
endomorphisms of a suitable vector bundle over a torus. The
fiberwise parts of these endomorphisms form a family of finite
dimensional quantizations, parameterized by the points of a
torus, which includes the quantization proposed in [HB].

\vfill\eject}
\section\intsec{Introduction}

\newsubsec
Quite distinct from its classical counterpart, there remains as yet no
well-accepted concept of quantum ergodicity. Several inequivalent
yet very natural approaches have been introduced. On the one
hand, a system is deemed ``quantum ergodic'' if it has a well-defined
classical limit which is itself ergodic [Z1,2], [S], [C], [EGI]. On
the other hand, the original notion of quantum ergodicity proposed
by von Neumann defines, roughly speaking, a system as quantum
ergodic if any observable is eventually distributed over the
eigenstates according to the weight of each eigenstate.

Let us discuss this latter notion first. Let $\cH$ be a Hilbert space,
$\fA$ a *-algebra of operators on $\cH$, and F a unitary quantum
evolution operator (called also the propagator). Then the quantum
system is ``quantum ergodic'' if for all observables $A\in \fA$,
and any $\varphi\in\cH$,
$$
\lim_{M\rightarrow\infty}{1\over M}\sum_{0\leq m\leq M-1}
(\varphi,F^mAF^{-m}\varphi)=\sum_{n=0}^{\infty}
|c_n|^{2}(\varphi_n,A\varphi_n),\ref{\vNerg}
$$
where $c_n$ is the $n$-th Fourier coefficient of a vector $\varphi$ with
respect to the eigenstates of $F$ spanning $\cH, \varphi=\sum_{n}c_n
\varphi_n$. Although the physical motivation behind this definition is
indeed appealing, it leads unfortunately to quite unexpected and
somewhat counterintuitive results. First of all, it applies only to
systems whose propagators have purely discrete spectra. Furthermore,
it can be readily shown that any system whose propagator spectrum is
simple (e.g. the one dimensional harmonic oscillator with generic
frequency) is, as a consequence of this definition, quantum ergodic.

In [S], [C], [Z1], a system is defined as quantum ergodic if the time
average (which is essentially the left hand side of \vNerg), smears
the quantum mechanics onto a ``classical limit state'' plus a quantum
mechanical correction which vanishes asymptotically in the classical
limit. The existence of such a state is a highly non-trivial result
and often a quantum system will not have such a ``classical limit''.
In [Z1], quantum ergodicity of a class of quantum dynamical systems,
called ``Gelfand-Segal systems'' are studied. By definition, a
Gelfand-Segal systems has a propagator whose spectrum is discrete.
This concept of quantum ergodicity seems to be particularly useful
in systems which arise as quantizations of the geodesic flow on a
compact manifold. For a discussion of quantized toral automorphisms
within this framework, see [Z2] and [BB].

\subsec
In this paper we study the ergodic properties of a class of quantum
dynamical systems whose spectra are continuous. The examples we
discuss are the quantized cat and Kronecker maps. We work within
the algebraic quantization scheme which emphasizes the role of
observables in quantum kinematics and dynamics. In the context of
toral automorphisms, such a scheme was discussed in [KL]. That paper
contains also an extensive list of references to the original
literature concerning quantized toral automorphisms and algebraic
quantization. A particularly natural and convenient choice of the
algebra of observables turns out to be the $\bC^*$-algebra $\fA_\hbar$
generated by Toeplitz operators $T_\hbar(f)$ on the Bargmann space with
$\bZ^2$-invariant symbols $f$. These Toeplitz operators are simply
anti-Wick ordered quantizations of classical observables. The two
properties which make Toeplitz quantization very natural are:
(i) Toeplitz quantization is positivity preserving,
$$
T_\hbar(f)\geq 0, \quad\hbox{if } f\geq 0,\ref{\toeplpositiv}
$$
and (ii) Toeplitz quantization is continuous in the symbol,
$$
\Vert T_\hbar(f)\Vert\leq\Vert f\Vert_\infty,\ref{\toeplcont}
$$
where $\Vert\cdot\Vert$ is the operator norm, and
where $\Vert\cdot\Vert_\infty$ is the sup norm. These properties
have important consequences for the study of the semiclassical
limit of the quantum system.

The results established in this paper have the form
$$
\lim_{M\rightarrow\infty}{1\over M}\sum_{0\leq m\leq M-1}
F^mAF^{-m}=\tau_\hbar(A)I,\ref{\inftimelim}
$$
where $A$ is an element of $\fA_\hbar$, and where $\tau_\hbar$ is
a trace on $\fA_\hbar$. This trace is invariant under the quantum
dynamics and reduces to the classical ensemble average in the
limit as $\hbar\rightarrow 0$. It can be thus thought of as the
quantum ensemble average. The limit in \inftimelim\ is in the
sense of weak topology.

\subsec
The paper is organized as follows. In section II we present the classical
maps, briefly review some of the results of [KL] relevant to this paper,
in particular the quantum time evolution operator in Bargmann space,
and introduce a trace $\tau_\hbar$ on the algebra of observables.
The quantization of the dynamics derived in [KL] together with the
concept of the trace enable us to show ergodicity and mixing of the cat
map in Section III. That is, (i) the time average of an observable
converges weakly in the large time limit to the trace of that
observable, and (ii) for observables A and B, $\tau_\hbar(F^MAF^{-M} B)$
converges in the large time limit to the product $\tau_\hbar(A)
\tau_\hbar(B)$. Related results, within a different quantization scheme,
had previously been discussed in [BNS]. In section IV we show that the
quantum Kronecker map is ergodic but not mixing. In section V, we
study the structure of the quantized cat maps. For the values of
Planck's constant satisfying the geometric quantization ``integrality
condition'', we construct an isomorphism between the Bargmann space
and the Hilbert space of sections of a vector bundle over the torus.
Under this isomorphism, the algebra automorphism defining the quantum
cat map becomes a unitary vector bundle endomorphism. This yields a
family of finite dimensional quantizations of the cat dynamics
parameterized by the points on the torus. A particular element of
this family reduces to the quantization scheme proposed originally
in [HB]. Section VI contains similar results for the quantized
Kronecker maps.

\section\torautsec{Quantized toral automorphisms}

\newsubsec
We begin here with a brief review of the systems we shall study.
We restrict ourselves to two of the simplest and well known maps
of the torus: Arnold's cat map and the Kronecker map. For a more
thorough treatment we refer the reader to [A], [AW], [CFS].

The cat map is a linear automorphism of the torus, with one step
classical evolution represented by an element $\gamma\in SL(2,\bZ)$,
$$
\gamma=\pmatrix{a&b\cr
                c&d\cr}.\ref{\gammadef}
$$
It can be readily verified that $|\tr(\gamma)|>2$ corresponds to
uniformly hyperbolic dynamics, while $|\tr(\gamma)|<2$ yields
elliptic motion. Since we are interested in chaotic dynamics, we
restrict ourselves to $|\tr(\gamma)|>2$. In this case, the dynamics
evolves locally along two linearly independent eigenvectors
which are not orthogonal. Indeed, the slopes are irrationally
related. The two eigenvalues of \gammadef\ $\mu_1$ and $\mu_2$
satisfy
$$
\mu_1\mu_2=1,\num
$$
with $|\mu_1|>1$ and $|\mu_2|<1$ corresponding to flow along
unstable and stable axes, respectively.

As in [KL], we write the dynamics
$$
(x_1,\;x_2)\longrightarrow(ax_1+bx_2,\;cx_1+dx_2)\num
$$
in terms of the complex variable $z=(x_1+ix_2)/\sqrt{2}$ via
$$
z\longrightarrow\ol\alpha z+\beta\ol z.\num
$$
The factor of $\sqrt{2}$ in the denominator serves to make the
transformation $(x_1,\;x_2)\rightarrow z$ a symplectomorphism. Note
that $\alpha$ and $\beta$ are simply the complex cat map parameters,
with
$$
\eqalign{
\alpha&=\big(a+d+i(b-c)\big)/2,\cr
\beta&=\big(a-d+i(b+c)\big)/2,\cr}\num
$$
and
$$
|\alpha|^2-|\beta|^2=1.\num
$$

The classical Kronecker map is an even simpler automorphism of the
torus defined by
$$
(x_1,x_2)\longrightarrow(x_1+\omega_1,x_2+\omega_2),
$$
(or equivalently $z\longrightarrow z+\omega$,) where the frequencies
$\omega_1$ and $\omega_2$ are linearly independent over
$\bZ$, i.e. $\omega_1/\omega_2$ is irrational. It is a well known
result, see e.g. [CFS], that the map is ergodic; however
because of its simple uniform motion, it is not mixing. Furthermore,
if we consider the same map with $\omega_1/\omega_2$ rational, then
the classical dynamics is no longer ergodic. Rather, it is described
by periodic orbits whose lengths are related to how closely $\omega_1/
\omega_2$ approximates an irrational number.

\subsec
We shall use the quantization presented in [KL]. For a full
account of the method the reader is referred to the original paper.
Here we only summarize the results.

We work in a Bargmann representation with a Hilbert space $\hil$
consisting of entire functions on $\bC$ which are square integrable
with respect to the measure $d\mu_\hbar(z)=(\pi\hbar)^{-1}\exp(-
|z|^2/\hbar)d^2z$. Quantizations of classical functions over phase
space (which are ``functions'' over the quantized phase space) generate
naturally a quantum mechanical algebra of observables. Of course, a
particular choice of quantization must be given. We choose, as in [KL],
the anti-Wick quantization, and define the algebra of observables to be
the $\bC^*$-algebra $\fA_\hbar$ generated by Toeplitz operators. A
Toeplitz operator with symbol $f$ is given by
$$
T_\hbar(f)\varphi(z)=\int_{\bC}e^{z\ol w/\hbar}f(w)\varphi(w)
d\mu_\hbar(w).\ref{\toeplitz}
$$
The Hilbert space $\hil$ carries a unitary projective representation
of the group of translations of $\bC$ given by
$$
U(\zeta)\varphi(z)=e^{(\ol\zeta
z-|\zeta|^2/2)/\hbar}\varphi(z-\zeta),\ref{\repres}
$$
with the property
$$
U(\zeta)U(\xi)=e^{i\im(\ol\zeta\xi)/\hbar}U(\zeta+\xi).\ref{\projrep}
$$

Consider the following operators:
$$
\eqalign{
U&=U(-i\hbar\pi\sqrt{2} ),\cr
V&=U(\hbar\pi\sqrt{2}).\cr}\ref{\generators}
$$
These operators are generators of the
algebra $\fA_\hbar$, and obey the commutation relation
$$
UV=e^{i\lambda}VU,\ref{\relation}
$$
where $\lambda=4\pi^2\hbar$.

Quantum cat dynamics in the Bargmann representation is
effected by the unitary operator
$$
F\varphi(z)=|\alpha|^{-1/2}\exp\big(-{{\ol\beta
z^2}\over{2\hbar\alpha}}
\big)\;\int_{\bC}\exp\big({{\ol wz}\over{\hbar\alpha}}+{{\beta\ol w^2}
\over{2\hbar\alpha}}\big)\varphi(w)\; d\mu_\hbar(w).\ref{\fdef}
$$
Indeed, it was shown explicitly that for $U$ and $V$ defined in
\generators ,
$$
\eqalign{
U^\prime&=FUF^{-1}=e^{-i\lambda ab/2}U^aV^b,\cr
V^\prime&=FVF^{-1}=e^{-i\lambda cd/2}U^cV^d.\cr}\ref{\dynamics}
$$
Furthermore, this $F$ has a well defined $\hbar\rightarrow 0$
limit yielding the desired classical dynamics.

For the Kronecker map, the unitary operator which effects the
dynamics is readily shown to be $K=U(-\omega)$:
$$
\eqalign{
KUK^{-1}&=e^{2\pi i\omega_1}U,\cr
KVK^{-1}&=e^{2\pi i\omega_2}V,\cr}\ref{\kronecker}
$$
where $\omega=(\omega_1+i\omega_2)/\sqrt{2}$.

\subsec
We also need a trace on the algebra $\alg$. Let $\varphi\in\hil$
be an arbitrary vector of norm one. For $S\in\alg$ we define
$$
\tau_\hbar(S)=\int_{\bT^2}(U(l)\varphi,SU(l)\varphi)d^2l.\ref{\trref}
$$
This functional has a number of remarkable properties which we
summarize in the theorems below.
\thm\trthm{The functional $\tau_\hbar$ has the following properties:
\item{$1^\circ$.} It is a state on $\alg$;
\item{$2^\circ$.} Its value is given by
$$
\tau_\hbar(U^mV^n)=\delta_{m0}\delta_{n0};\ref{\trvalref}
$$
\item{$3^\circ$.} It is a trace on $\alg$.
}

In particular, $\tau_\hbar$ is independent of the choice of $\varphi$
and from \trvalref\ coincides with the standard trace on the quantized
torus.
\proof $1^{\circ}$. Indeed, $\tau_\hbar$ is continuous,
$$
|\tau_\hbar(S)|\leq\int_{\bT^2}\|S\|\,\|U(l)\varphi\|^2d^2l
=\|S\|,
$$
positive,
$$
\tau_\hbar(S^\dagger S)=\int_{\bT^2}\|SU(l)\varphi\|^2d^2l\geq 0,
$$
and normalized,
$$
\tau_\hbar(I)=\int_{\bT^2}(U(l)\varphi,U(l)\varphi)d^2l
=\int_{\bT^2}\|\varphi\|^2d^2l=1.
$$
\noindent
$2^\circ$. This is a direct calculation:
$$
\eqalign{
\tau_\hbar(U^mV^n)&=\int_{\bT^2}(\varphi,U(-l)U(-im\pi\hbar\sqrt{2})
U(n\pi\hbar\sqrt{2})U(l)\varphi)d^2l\cr
&=\int_{\bT^2}e^{i\pi\sqrt{2}(m\re l + n\im l)}(\varphi,U(-l-im\pi\hbar
\sqrt{2})U(l+n\pi\hbar\sqrt{2})\varphi)d^2l\cr
&=\int_{\bT^2}e^{2i\pi\sqrt{2}(m\re l + n\im l)+2i\pi^2\hbar mn}
(\varphi,U((n-im)\pi\hbar\sqrt{2})\varphi)d^2l\cr
&=\int_0^1\int_0^1e^{2i\pi(mx + ny)+2i\pi^2\hbar mn}dxdy\;
(\varphi,U((n-im)\pi\hbar\sqrt{2})\varphi)=\delta_{m0}\delta_{n0}.\cr
}
$$
\noindent
$3^\circ$. This follows from $2^\circ$. $\square$

\noindent
{\bf Remark.} Notice that if $\varphi\in\hil$ is chosen to be the
ground state $\varphi_0=1$, then
$$
\varphi_l(z):=U(l)\varphi_0(z)=e^{\ol lz/\hbar-|l|^2/2\hbar}=
e^{-|l|^2/2\hbar}\sum_{n=0}^\infty{1\over{n!}}\big({\ol l}A^{\dagger}
/{\hbar}\big)^n\varphi_0(z),\ref{\coherstate}
$$
where $A^\dagger$ is the creation operator. Thus the trace of the
operator $S$ is the sum over the coherent states basis representation
restricted to the fundamental domain of the expectation value of $S$.

Another interesting fact about $\tau_\hbar$ is that its value on
a Toeplitz operator is equal to the integral of the symbol of that
operator. This is quite remarkable in that with our choice of
quantization, the quantum expectation value yields exactly the
classical value independent of Planck's constant.
\thm\treqthm{For any symbol $f\in\ct$,
$$
\tau_\hbar((T_\hbar(f))=\tau(f).\ref{\trequref}
$$
where $\tau(f)$ is the phase-space integral of f over the torus.}
\proof It is sufficient to prove this for $f$ a pure harmonic.
The general case will follow by linearity and continuity. Let
$$
\eqalign
{
f_\zeta(z)&:=e^{2\pi^2\hbar|\zeta|^2}e^{2\pi(\ol \zeta z-\zeta\ol z)} \cr
&=e^{\pi^2\hbar(m^2+n^2)}e^{2\pi i (nx+my)},\cr
}
\ref{\harmonicref}
$$
where $\zeta=(m-in)/\sqrt{2}$, $m,n\in\bZ$. Then, using \toeplitz, we find
$$
T_\hbar(f_\zeta)=U^nV^me^{2\pi^2i\hbar mn}
\ref{\toepurharm}.
$$
Using part $2^\circ$ of \trthm , we conclude that
$$
\tau_\hbar(T_\hbar(f_\zeta))=\delta_{m0}\delta_{n0}=
\int_{\bT^2}f_\zeta(z)d^2z,
$$
as claimed. $\square$

\section\catsec{Ergodic properties of the quantized cat map}

\newsubsec
In this section we study the ergodic properties of the quantized cat
map. We prove that the dynamics generated by this map has a property
which is a quantum mechanical analog of the strong mixing property.
Furthermore, we show that the quantized cat dynamics is ergodic in the
sense that the time average of an observable tends to its ensemble
average given by the trace $\tau_\hbar$.

\thm{\mixingthm}{{\rm (Strong mixing)} For any $A,B\in\alg$,
$$
\lim_{M\rightarrow\infty}\tau_\hbar(F^MAF^{-M}B)=
\tau_\hbar(A)\tau_\hbar(B).\ref{\mixingref}
$$}
\proof
We proceed in two steps.

\noindent
{\it Step 1.} We assume first that $A=T_\hbar(f_\zeta),\;B=T_\hbar
(f_\eta)$, with $f_\zeta,\;f_\eta$ of the form {\harmonicref}. We have
to show that the limit in {\mixingref} is $1$, if $\zeta=\eta=0$,
and 0, otherwise. A direct calculation (see Section III of [KL])
shows that
$$
FT_\hbar(f_\zeta)F^{-1}=T_\hbar(f_{\gamma^{-1}\zeta}),\ref{\evoloneref}
$$
and consequently
$$
F^mT_\hbar(f_\zeta)F^{-m}=T_\hbar(f_{\gamma^{-m}\zeta}).\ref{\evolmref}
$$
Furthermore, as a consequence of {\toepurharm},
$$
T_\hbar(f_\zeta)T_\hbar(f_\eta)=\epsilon(\zeta,\eta)T_\hbar(f_{\zeta+\eta}),
\ref{\multipref}
$$
where $\epsilon(\zeta,\eta)$ is such that
$$
|\epsilon(\zeta,\eta)|=1,\qquad\epsilon(0,\eta)=\epsilon(\zeta,0)=1.
\num
$$
As a consequence,
$$
\eqalign{
\tau_\hbar(F^MT_\hbar(f_\zeta)F^{-M}T_\hbar(f_\eta))&=\epsilon
(\gamma^{-M}\zeta,\eta)\tau_\hbar(T_\hbar(f_{\gamma^{-M}\zeta+\eta}))\cr
&=\epsilon(\gamma^{-M}\zeta,\eta)\int_{\bT^2}f_{\gamma^{-M}\zeta+\eta}
(z)d^2z.\cr}
$$
If $\zeta=\eta=0$, then the above expression is equal to $1$. For
$\zeta=0,\;\eta\neq 0$, $\int_{\bT^2}f_\eta(z)d^2z=0$. Let $\zeta
\neq 0,\;\eta\neq 0$. Since $\gamma$ is hyperbolic, there is $M_0$
such that for all $M\geq M_0$, $\gamma^{-M}\zeta+\eta\neq 0$, and
thus $\int_{\bT^2}f_{\gamma^{-M}\zeta+\eta}(z)d^2z=0$, for all $M\geq M_0$.

\noindent
{\it Step 2.} As a consequence of {\it Step 1}, {\mixingref} holds
for any $A=A_0:=T_\hbar(f)$ and $B=B_0:=T_\hbar(g)$, where $f$ and
$g$ are finite linear combinations of simple harmonics. Any element
of $\alg$ is a norm limit of such operators. Using the continuity
of $\tau_\hbar$ and unitarity of $F$ we obtain the inequality
$$
\eqalign{
|\tau_\hbar(F^MAF^{-M}B)-&\tau_\hbar(A)\tau_\hbar(B)|\cr
&\leq|\tau_\hbar(F^MA_0F^{-M}B_0)-\tau_\hbar(A_0)\tau_\hbar(B_0)|\cr
&\quad+|\tau_\hbar(F^MA_0F^{-M}(B-B_0))|+|\tau_\hbar(A_0)||\tau_\hbar
(B-B_0)|\cr
&\quad+|\tau_\hbar(F^M(A-A_0)F^{-M}B_0)|+|\tau_\hbar(A-A_0)||\tau_\hbar
(B_0)|\cr
&\quad+|\tau_\hbar(F^M(A-A_0)F^{-M}(B-B_0))|+|\tau_\hbar(A-A_0)|
|\tau_\hbar(B-B_0)|\cr
&\leq|\tau_\hbar(F^MA_0F^{-M}B_0)-\tau_\hbar(A_0)\tau_\hbar(B_0)|\cr
&\quad+2(\Vert A_0\Vert\Vert B-B_0\Vert+\Vert A-A_0\Vert\Vert B_0\Vert
+\Vert A-A_0\Vert\Vert B-B_0\Vert),\cr}
$$
from which {\mixingref} follows. $\square$

As a corollary, we obtain the following mixing property for observables
which are Toeplitz operators.
\cor\weakmixing{For $f,g\in C(\bT^2)$,
$$
\lim_{m\rightarrow\infty}\tau_\hbar\big(F^mT_\hbar(f)F^{-m}
T_\hbar(g)\big)=\tau(f)\tau(g).\num
$$}
\proof This is a consequence of \trequref . $\square$

\subsec
Now we formulate the ergodic theorem for the quantized cat
dynamics. For an operator $S$, define its time average over a period
of time $M$:
$$
\langle S\rangle_M:={1\over M}\sum_{0\leq m\leq M-1}F^mSF^{-m}.
\ref{\timeavref}
$$
The theorem below asserts that for any $A\in\alg$, the sequence
$\langle A\rangle_M$ converges to $\tau_\hbar(A)I$ in the weak
operator topology.
\thm{\catergthm}{{\rm (Ergodicity of the quantized cat map)} For
any $A\in\fA_\hbar$, and $\varphi,\psi\in\hil$,
$$
\lim_{M\rightarrow\infty}(\varphi,\langle A\rangle_M\psi)=
\tau_\hbar(A)(\varphi,\psi).\ref{\catergref}
$$}
\proof
We proceed in three steps.

\noindent
{\it Step 1.} Let $A=T_\hbar(f_\zeta)$, where $f_\zeta$ is a simple
harmonic. If $\zeta=0$, then $\langle A\rangle_M=I$, and {\catergref}
holds trivially. Let $\zeta\neq 0$; we have to show that
$$
\lim_{M\rightarrow\infty}(\varphi,\langle T_\hbar(f_\zeta)
\rangle_M\psi)=0.\ref{\auxilergref}
$$
Assume now that $\varphi$ and $\psi$ are normalized coherent states
of the form {\coherstate}, $\varphi=\varphi_\xi,\;\psi=\varphi_\eta$.
Then
$$
\eqalign{
(\varphi_\xi,T_\hbar(f_\zeta)\varphi_\eta)&=\int_{\bC}
\ol{\varphi_\xi(z)}f_\zeta(z)\varphi_\eta(z)d\mu_\hbar(z)\cr
&=e^{-(|\xi|^2-2\xi\ol\eta+|\eta|^2)/2\hbar-2\pi^2\hbar
|\zeta|^2+2\pi(\ol\zeta\xi-\zeta\ol\eta)},\cr}
$$
and so
$$
|(\varphi_\xi,T_\hbar(f_\zeta)\varphi_\eta)|\leq
e^{-2\pi^2\hbar|\zeta|^2+2\pi(|\xi|+|\eta|)|\zeta|}.
$$
Consequently,
$$
\eqalign{
|(\varphi_\xi,\langle T_\hbar(f_\zeta)\rangle_M\varphi_\eta)|
&\leq{1\over M}\sum_{0\leq m\leq M-1}e^{-2\pi^2\hbar|\gamma^{-m}
\zeta|^2+2\pi(|\xi|+|\eta|)|\gamma^{-m}\zeta|}\cr
&\leq{1\over M}\sum_{0\leq m\leq M-1}e^{-O(1)|\mu_1|^{2m}}
\longrightarrow 0,\cr}
$$
where $\mu_1$ is the eigenvalue of $\gamma$ with $|\mu_1|>1$.

\noindent
{\it Step 2.} By means of {\it Step 1}, {\auxilergref} holds for $\varphi=
\varphi_0$ and $\psi=\psi_0$ which are finite linear combinations
of coherent states. Any vector in $\hil$ is a norm limit of
such elements. Observe also that the time average of an operator
obeys the following inequality:
$$
\Vert\langle S\rangle_M\Vert\leq\Vert S\Vert.\ref{\contref}
$$
This leads to the inequality
$$
\eqalign{
|(\varphi,\langle T_\hbar(f_\zeta)\rangle_M\psi)|&\leq
|(\varphi_0,\langle T_\hbar(f_\zeta)\rangle_M\psi_0)|\cr
&\quad+\Vert T_\hbar(f_\zeta)\Vert(\Vert\varphi-\varphi_0\Vert
\Vert\psi_0\Vert+\Vert\varphi_0\Vert\Vert\psi-\psi_0\Vert+
\Vert\varphi-\varphi_0\Vert\Vert\psi-\psi_0\Vert),\cr}
$$
which yields {\auxilergref}.

\noindent
{\it Step 3.} As a consequence of {\it Step 2}, {\catergref}
holds for any $A=A_0:=T_\hbar(f)$, where $f$ is a finite linear
combination of simple harmonics. Any element of $\alg$ is a norm
limit of such operators. Using the continuity of $\tau_\hbar$ and
{\contref} we obtain the inequality:
$$
|(\varphi,\langle A\rangle_M\psi)-\tau_\hbar(A)(\varphi,\psi)|
\leq|(\varphi,\langle A_0\rangle_M\psi)-\tau_\hbar(A_0)(\varphi,
\psi)|+2\Vert A-A_0\Vert\Vert\varphi\Vert\Vert\psi\Vert,
$$
and our claim follows. $\square$

A simple corollary to the above theorem is the following result.
It states that the time average of a quantum observable $T_\hbar(f)$
(namely the anti-Wick quantization of the classical observable
$f$) converges weakly to the ensemble average of the classical
observable $f$.
\cor\weakerg{For $f\in C(\bT^2)$, and $\varphi,\psi\in\hil$,
$$
\lim_{M\rightarrow\infty}(\varphi,\langle T_\hbar(f)\rangle_M\psi)
=\tau(f)(\varphi,\psi).\num
$$}
\section\kroneckersec{Ergodic properties of the quantized Kronecker
map}

We turn now to ergodicity of the Kronecker map. In this section,
we let $\langle\;\cdot\;\rangle_M$ denote the time average defined by
\timeavref, with $F$ replaced by $K$. First, we prove the ergodic
theorem for the quantized Kronecker dynamics. It states that the
time averages of an observable converge in norm to its ensemble
average (this is a somewhat stronger property than {\catergthm}
which involves weak convergence of time averages).
\thm\kronergthm{{\rm (Ergodicity of the quantized Kronecker map)} For
$A\in\fA_\hbar$,
$$
\lim_{M\rightarrow\infty}\|\langle A\rangle_M-\tau_\hbar(A)I
\|=0.\ref{\kronerg}
$$
In particular, for $f\in\ct$,
$$
\lim_{m\rightarrow\infty}\|\langle T_\hbar(f)\rangle_M-\tau(f)I
\|=0.\ref{\kronergtwo}
$$}
\proof
Let $f=f_\zeta$, where $f_\zeta$ is given by \harmonicref,
with the corresponding Toeplitz operator $T_\hbar(f_\zeta)$. Then
$$
K^mT_\hbar(f_\zeta)K^{-m}=e^{m\pi\sqrt{2}(\zeta\ol\omega-\ol\zeta\omega)}
T_\hbar(f_\zeta),\ref{\quantkronact}
$$
and so
$$
\langle T_\hbar(f_\zeta)\rangle = {1\over M}\sum_{0\leq m\leq M-1}
e^{m\pi\sqrt{2}(\zeta\ol\omega-\ol\zeta\omega)}T_\hbar(f_\zeta).
$$
This is equal to $1$ if $\zeta = 0$, while for $\zeta\neq 0$,
$$
\|\langle T_\hbar(f_\zeta)\rangle \|\leq O(1)\|T(f_\zeta)\|/M
=O(1)/M,
$$
uniformly in $\zeta$.

Simple continuity arguments similar to {\it Step 3} in the proof
of {\catergthm} conclude the proof of {\kronerg}. $\square$

True to its classical origins, while the quantum Kronecker map is
ergodic, it is not mixing.
\thm\kronmixthm{The quantized Kronecker map is not mixing in the
sense of {\mixingref}.}
\proof We construct a counter-example.  We take A to be the
Toeplitz operator for the pure harmonic, $A=T_\hbar(f_\zeta)$, where
$\zeta\neq0$, and $B=T_\hbar(f_{-\zeta})$.
Then, by means of {\quantkronact}, {\trequref}, and {\multipref},
$$
\eqalign{
|\tau_\hbar(K^mAK^{-m}B)-\tau_\hbar(A)\tau_\hbar(B)|
&=|\tau_\hbar(K^mT_\hbar(f_\zeta)
K^{-m}T_\hbar(f_{-\zeta}))-\tau(f_\zeta)\tau(f_{-\zeta})|\cr
&=|\tau_\hbar(e^{m\pi \sqrt{2}(\zeta \ol w -\ol\zeta
w)}T_\hbar(f_\zeta)T_\hbar(f_{-\zeta}))|\cr
&=1,\cr
}
$$
uniformly in $m$. Thus the Kronecker map does not satisfy the mixing
condition. $\square$

\section\catstructsec{The structure of the quantized cat map}

\newsubsec
The quantization method used in [KL] and in the present paper is
convenient to study global properties of the quantized cat dynamics.
For numerical analysis and detailed spectral properties of the
dynamics, the previous quantization schemes ([HB], [D]) seem more
suitable. In this section we establish a connection between our
quantization method of the cat dynamics to the previous methods.
Indeed, we will see that our scheme yields a continuum of quantizations
(similar to those of [HB]) parameterized by a point $\theta$ on the
two-dimensional torus. In the special case of $\theta=0$ and $\gamma$
of the structure
$$
\pmatrix{odd&even\cr
         even&odd}  \, \, \hbox {or }  \,\,  \pmatrix{even&odd\cr
                                  odd&even},\ref{\checkerboard}
$$
which was referred to as ``quantizable'' in [HB], we reproduce exactly
the result of [HB]. In the general case, the angle variable $\theta$
parametrizes the different quantizations that result from having a
classical phase space with a non-trivial topology. As in these earlier
works, we restrict Planck's constant to satisfy the integrality condition
$$
\hbar=1/2\pi N.\ref{\planck}
$$

We introduce the following notation:
$$
X:=U(-i/\sqrt{2}),\qquad Y:=U(1/\sqrt{2}),\ref{\xydef}
$$
and observe that as a consequence of \relation ,
$$
[X,Y]=0.\ref{\xycomm}
$$
The operators $X$ and $Y$ generate an action of the group $\bZ^2$
on $\hil$. We also verify easily that,
$$
\eqalign{
&[X,U]=0,\qquad [X,V]=0,\cr
&[Y,U]=0,\qquad [Y,V]=0,\cr}\num
$$
and so $X$ and $Y$ are in the commutant of $\fA_\hbar$. Finally, we
note that
$$
X=U^N,\qquad Y=V^N.\ref{\utonref}
$$

\subsec
We shall call a holomorphic function $\phi$ on $\bC$  a
$\bZ^2$-{\it automorphic form} if
$$
\eqalign{
X\phi(z)&=e^{2\pi i\theta_1}\phi(z),\cr
Y\phi(z)&=e^{2\pi i\theta_2}\phi(z),\cr}\num
$$
where $\theta=(\theta_1,\theta_2)\in\bT^2$. In other words,
$\bZ^2$-automorphic forms are simultaneous generalized eigenvectors of
$X$ and $Y$. Let $\cH_\hbar(\theta)$ denote the space of all
$\bZ^2$-automorphic forms with fixed $\theta$. Clearly,
$\phi\in\cH_\hbar(\theta)$ is uniquely determined once defined on the
fundamental domain $D=[0,1]\times[0,1]\subset\bR^2$. The space
$\cH_\hbar(\theta)$ has a natural inner product defined as an integral
over this domain:
$$
(\phi_1,\phi_2)=\int_D\ol{\phi_1(z)}\phi_2(z) d\mu_\hbar(z).
\ref{\innprod}
$$
(Note a similar integral over the entire complex plane {\it does not
converge}, hence the $\bZ^2$-automorphic forms are not in
$ \cH^2(\bC,d\mu_\hbar)$.) This inner product is a $\bZ^2$ version of
the familiar Petersson inner product. In the following lemma we
construct a natural orthonormal basis for the space $\cH_\hbar(\theta)$.
\lemma\basislemma{
\item{(i)} {The following functions are elements of $\cH_\hbar(\theta)$:
$$
\phi_m^{(\theta)}(z)=\;C_m(\theta)e^{-N\pi z^2+2\sqrt{2}\pi(\theta_1+m)z}
\sum_{k\in\bZ}e^{-N\pi k^2-2\pi(\theta_1+i\theta_2+m)k
+2\sqrt{2}N\pi kz},\ref{\basisdef}
$$
where
$$
C_m(\theta):=(2/N)^{1/4}e^{-\pi(\theta_1+m)^2/N-2\pi i\theta_2m/N}.
\ref{\normconst}
$$
They are periodic in $m$,
$$
\phi_{m+N}^{(\theta)}=\phi_m^{(\theta)},\ref{\mperiodicity}
$$
and furthermore,
$$
\phi_0^{(\theta)},\ldots,\phi_{N-1}^{(\theta)}\ref{\basis}
$$
are orthonormal vectors in $\cH_\hbar(\theta)$.}
\item{(ii)} {The space $\cH_\hbar(\theta)$ has dimension $N$. Consequently,
the functions \basis\ form an orthonormal basis for $\cH_\hbar(\theta)$.}
\item{\phantom{(ii)}}}

\noindent{\bf Remark.} We observe that our basis functions \basisdef\
can be written in terms of the Jacobi $\vartheta$ functions (see for
example [M]):
$$
\phi_m^{(\theta)}(z)=C_m(\theta)e^{-N\pi z^2 +2\sqrt{2}\pi(\theta_1 +m)z}
\vartheta(-i\sqrt{2}Nz +i(\theta_1 + i\theta_2 +m), iN),
$$
where
$$
\vartheta(\omega,\tau)=\sum_{k\in\bZ}e^{i\pi k^2 \tau +2\pi i
k\omega}.
$$
Expressions similar to {\basisdef} have been used before, see e.g. [LV],
and references therein.

\proof ({\it i}) Usual arguments show that \basisdef\ converges
on compact subsets of $\bC$ and thus defines an entire function. It
can be readily checked that $X\phi_m^{(\theta)}(z)=e^{2\pi i\theta_1}
\phi_m^{(\theta)}(z)$, $Y\phi_m^{(\theta)}(z)=e^{2\pi i\theta_2}
\phi_m^{(\theta)}(z)$, and so $\phi_m^{(\theta)}\in\cH_\hbar(\theta)$.
The periodicity condition \mperiodicity\ can be verified easily.

\noindent
To show that $\phi_m^{(\theta)}$, $0\leq m\leq N-1$, form an
orthonormal set, we compute
$$
\eqalign{
(\phi_m^{(\theta)},\phi_n^{(\theta)})=\;
&\ol{C_m(\theta)}C_n(\theta) \sum_{k,l\in\bZ}e^{-N\pi(k^2+l^2)
-2\pi(\theta_1-i\theta_2)k-2\pi(\theta_1+i\theta_2)l
-2\pi(mk+nl)}\cr
&\times N\int_De^{-N\pi|z+\ol z|^2+2\sqrt{2}\pi(\theta_1+m+Nk)
\ol{z}+2\sqrt{2}\pi(\theta_1+n+Nl)z}d^2z\cr
=&\ol{C_m(\theta)}C_n(\theta)\sum_{k,l\in\bZ}e^{-N\pi(k^2+l^2)
-2\pi(\theta_1-i\theta_2)k-2\pi(\theta_1+i\theta_2)l
-2\pi(mk+nl)}\cr
&\times N\int_0^1e^{-2N\pi x^2+2\pi(2\theta_1+m+n+N(k+l))x}dx
\int_0^1e^{-2\pi i(m-n+N(k-l))y}dy.\cr}
$$
Let $m\neq n$. Since both $m$ and $n$ are between $0$ and $N-1$,
the expression $m-n+N(k-l)$ does not vanish and so $(\phi_m^{(\theta)},
\phi_n^{(\theta)})=0$. Let $m=n$. Then
$$
\eqalign{
(\phi_m^{(\theta)},\phi_m^{(\theta)})=\;&N|C_m(\theta)|^2\sum_{k\in\bZ}
\;e^{-2N\pi k^2-4\pi(\theta_1+m)k}\int_0^1e^{-2N\pi x^2+
4\pi(\theta_1+m+Nk)}dx\cr
=\;&N|C_m(\theta)|^2\sum_{k\in\bZ}\int_0^1e^{-2N\pi(x-k)^2+
4\pi(\theta_1+m)(x-k)}\cr
=\;&N|C_m(\theta)|^2\int_{\bR}e^{-2N\pi x^2+4\pi(\theta_1+m)}dx\cr
=\;&N|C_m(\theta)|^2(2N)^{-1/2}e^{2\pi(\theta_1+m)^2/N}\cr
=\;&1,}
$$
and the claim is proved.

{\it (ii)} We proceed in steps.

\noindent
{\it Step 1.} We shall first show that
$\phi\in\cH_\hbar(\theta)$, when
considered as a function of $z$, has exactly $N$ zeros inside any
fundamental domain. Observe that $\phi\in\cH_\hbar(\theta)$ satisfies
$$
\eqalign{
\phi(z-1/\sqrt{2})&=e^{-\sqrt{2}\pi Nz + {N\pi/2}\pi + 2\pi i\theta_2}
\phi(z),\cr
\phi(z+i/\sqrt{2})&=e^{-\sqrt{2}\pi Niz + {N\pi/2} + 2\pi i\theta_1}
\phi(z).\cr}\ref{\periodrel}
$$
Using the argument principle of elementary complex analysis and
\periodrel , we readily see that $\phi$ has precisely $N$ zeros inside
a fundamental domain.

\noindent
{\it Step 2.} For a torus, the Riemann-Roch theorem [FK] can be stated
as follows. For any divisor $D$,
$$
r(D^{-1})={\rm deg}D+i(D).\ref{\riemann}
$$
Recall that a divisor $D=P_1^{n_1}\ldots P_k^{n_k}$ is the collection
of points $P_1,\ldots,P_k$ on the torus, with integers $n_1,\ldots,n_k$
assigned to each point. Note that implicitly we assign to all other
points $n=0$. The inverse $D^{-1}$ of the divisor $D$ is simply
$D^{-1}=P_1^{-n_1}\ldots P_k^{-n_k}$. An example of a divisor is the
set of zeros ($n_j>0$) and poles ($n_j<0$) of a meromorphic
function $f$. We denote such a divisor by $(f)$. Similarly, given a
meromorphic $1$-form $\omega$ we denote its divisor by $(\omega)$.
An order relation among divisors can be defined as follows: for $D_1=
P_1^{n_1}\ldots P_k^{n_k}$, $D_2=P_1^{m_1}\ldots P_k^{m_k}$,
$D_1\geq D_2$, if $n_j\geq m_j$, for all $j$. The degree of a divisor
${\rm deg}D$ is defined by ${\rm deg}D=\sum_jn_j$. We set $r(D)$ equal
to the dimension of the vector space $L(D)$ of meromorphic functions
$f$ such that $(f)\geq D$. Likewise, $i(D)$ is the dimension of the
space of meromorphic one-forms $\omega$ such that $(\omega)\geq D$.

We now take $D$ to be the zero divisor of $\phi$. As a consequence of
{\it Step 1}, ${\rm deg}D=N$. Furthermore [FK], for  ${\rm deg}D>0$,
$i(D)=0$. Thus we see that
$$
r(D^{-1})=N.\ref{\riemanntwo}
$$

\noindent
{\it Step 3.} Notice that for $\phi\in\cH_\hbar(\theta)$ and
$\phi_m^{(\theta)}$ given by \basisdef , the quotients
$$
\psi_m(z):=\phi_m^{(\theta)}(z)/\phi(z),\ref{\quotients}
$$
define meromorphic functions on the torus. By means of part {\it (i)}
of the theorem, they are linearly independent. Since by construction
$(\psi_m)\geq D^{-1}$, the result of {\it Step 2} implies that the set
$\{\psi_0,\ldots,\psi_{N-1}\}$ spans $L(D^{-1})$. Suppose now that
$\phi$ is linearly independent of $\phi_0,\ldots,\phi_{N-1}$. It
follows that $1,\psi_0, \ldots,\psi_{N-1}$ are linearly independent,
which contradicts \riemanntwo . $\square$

\noindent
{\bf Remark.} The space $\dint_{\bT^2}\cH_\hbar(\theta)d\theta$ is thus
the Hilbert space of square integrable sections of an $N$-dimensional
vector bundle $Q\bT^2_\hbar\rightarrow\bT^2$. The bundle $Q\bT^2_\hbar$
does not admit a global continuous section and is thus topologically
non-trivial.

\subsec
In the next lemma, we construct an isomorphism between $\hil$ and
the direct integral of the spaces $\cH_\hbar(\theta)$. Under this
isomorphism, the actions of $U$ and $V$ are block diagonal. Let
$\phi_m\in\dint_{\bT^2}\cH_\hbar(\theta)d\theta$ be a (discontinuous
in $\theta$) element defined by $\phi_m(\theta,z)=\phi^{(\theta)}_m(z),\;
0\leq\theta_j<1$.

\lemma\isomlemma{There is an isomorphism
$$
\kappa:\;\cH^2(\bC,d\mu_\hbar)\longrightarrow\dint_{\bT^2}
\cH_\hbar(\theta)\; d\theta,\num
$$
such that
$$
\eqalign{
\kappa U\kappa^{-1}\phi_m(\theta,z)&=e^{2\pi i(\theta_1+m)/N}
\phi_m(\theta,z),\cr
\kappa V\kappa^{-1}\phi_m(\theta,z)&=e^{2\pi i\theta_2/N}
\phi_{m+1}(\theta,z).\cr}\ref{\uvaction}
$$}

\proof We set for $\varphi\in\hil$,
$$
\kappa(\varphi)(\theta,z):=\sum_{m,n\in\bZ^2}X^mY^n\varphi(z)
e^{-2\pi im\theta_1-2\pi in\theta_2},\ref{\kappadef}
$$
and verify easily that $\kappa(\varphi)(\theta,\;\cdot)\in
\cH_\hbar(\theta)$. For $\psi\in\dint_{\bT^2}\cH_\hbar(\theta)
d\theta$ we set
$$
\kappa^{-1}(\psi)(z):=\int_{\bT^2}\psi(\theta,z)d\theta,\ref{\kappaindef}
$$
and note that $\kappa^{-1}(\psi)\in\hil$. We verify that $\kappa$ and
$\kappa^{-1}$ are inverse to each other. Indeed, for $\varphi\in\hil$,
$$
\eqalign{
(\kappa^{-1}\kappa(\varphi))(z)
&=\int_{\bT^2}\kappa(\varphi)(\theta,z)d\theta\cr
&=\int_{\bT^2}\sum_{m,n\in\bZ}X^mY^n\varphi(z)
e^{-2\pi im\theta_1-2\pi in\theta_2}d\theta\cr
&=\;\varphi(z).\cr}
$$
Now, let $\phi\in\dint_{\bT^2}\cH_\hbar(\theta)d\theta$. Expanding
in a Fourier series, we have
$$
\phi(\theta,z)=\sum_{m,n\in\bZ}\widehat\phi_{m,n}(z)
e^{-2\pi im\theta_1-2\pi in\theta_2},
$$
where $\widehat\phi_{m,n}(z)=\int_{\bT^2}\phi(\theta,z)
e^{2\pi im\theta_1+2\pi in\theta_2}\;d\theta$. Consequently,
$$
\eqalign{
(\kappa\kappa^{-1}(\phi))(\theta,z)
&=\sum_{m,n\in\bZ}X^mY^n\kappa^{-1}(\phi)(z)
\;e^{-2\pi im\theta_1-2\pi in\theta_2}\cr
&=\sum_{m,n\in\bZ}X^mY^n\int_{\bT^2}\phi(\eta,z)d\eta
\;\;e^{-2\pi im\theta_1-2\pi in\theta_2}\cr
&=\sum_{m,n\in\bZ}\int_{\bT^2}\phi(\eta,z)
e^{2\pi im\eta_1+2\pi in\eta_2}d\eta
\;\;e^{-2\pi im\theta_1-2\pi in\theta_2}\cr
&=\;\phi(\theta,z),\cr}
$$
proving that $\kappa$ is an isomorphism.

We will prove the first of the identities \uvaction\ only; the
proof of the second one is identical. The calculation goes as
follows:
$$
\eqalign{
U\kappa^{-1}\phi_m(z)&=\int_{\bT^2}C_m(\theta)e^{2N\pi(iz/N\sqrt{2}-
1/4N^2)-N\pi(z+i/N\sqrt{2})^2+2\sqrt{2}\pi(\theta_1+m)(z+i/N\sqrt{2})}\cr
&\qquad\times\sum_{k\in\bZ}e^{-N\pi k^2-2\pi(\theta_1+i\theta_2+m)k+
2\sqrt{2}N\pi k(z+i/N\sqrt{2})}\;d\theta\cr
&=(2N)^{-1/4}\int_{\bT^2}e^{-\pi(\theta_1+m)^2/N-2\pi i\theta_2m/N
-N\pi z^2+2\sqrt{2}\pi(\theta_1+m)z+2\pi i(\theta_1+m)/N}\cr
&\qquad\times\sum_{k\in\bZ}e^{-N\pi k^2-2\pi(\theta_1+i\theta_2
+m)k+2\sqrt{2}N\pi kz}\;d\theta\cr
&=\int_{\bT^2}e^{2\pi i(\theta_1+m)/N}\phi_m(\theta,z)\;d\theta .\cr}
$$
Hence $\kappa U\kappa^{-1}\phi_m(\theta,z)=e^{2\pi i(\theta_1+m)/N}
\phi_m(\theta,z)$. $\square$

\subsec
We now wish to reformulate the quantum cat dynamics on the direct
integral space $\int_{\bT^2}^\oplus\cH_\hbar(\theta)d\theta$.
We begin with the following lemma, which demonstrates that the
quantum evolution operator $F$ yields a smooth endomorphism of
the bundle $Q\bT^2_\hbar$. Recall that $F$ is the integral operator
given by \fdef .
\lemma\koopman{For $\phi\in\cH_\hbar(\theta)$,
$$
XF\phi(z)=e^{2\pi i (\gamma^{-1}\theta
+\Delta_{\gamma^{-1}})_{1}} F\phi(z)
$$
$$YF\phi(z)=e^{2\pi i (\gamma^{-1}\theta
+\Delta_{\gamma^{-1}})_{2}} F\phi(z),
$$
where
$$
\Delta_\gamma=(Nab/2,Ncd/2).\ref{\deltadef}
$$
Consequently, $F$ maps unitarily $\cH_\hbar(\theta)$ onto
$\cH_\hbar(\gamma^{-1}\theta+\Delta_{\gamma^{-1}})$.}

\proof Using \fdef, \xydef, and \projrep, we compute
$$
\eqalign{
XF\phi(z)=&|\alpha|^{-1/2}e^{2\pi N(iz/\sqrt{2}-1/4)+\pi
N\ol\beta (z + i/\sqrt{2})^2/\alpha}\cr
&\times N\int_{\bC}e^{2 \pi N\ol w(z+i/\sqrt{2})/\alpha+
\pi N\beta\ol w^2/\alpha - 2 \pi N |w|^2}
\phi(w)\; d^2w.}
$$
Making the change of variables $w^\prime = w - 2^{-1/2}(b + id)$ and
using $X\phi(z)=e^{2\pi i\theta_1}\phi(z)$ and
$Y\phi(z)=e^{2\pi i\theta_2}\phi(z)$ we may reduce
this, after some straightforward algebra, to
$$
\eqalign{
XF\phi(z)=&e^{-i \pi Nbd + 2 \pi i(\theta_1d-\theta_2b)}
F\phi(z)\cr
=&e^{2\pi i(\gamma^{-1}\theta + \Delta_{\gamma^{-1}})_1}F\phi(z).
}
$$
Similarly,
$$
YF\phi(z)=e^{2\pi i(\gamma^{-1}\theta +
\Delta_{\gamma^{-1}})_2}F\phi(z),
$$
and the lemma is thus proved. $\square$

An equivalent way of stating the above lemma is in the form of the
following commutation relations:
$$
\eqalign{
&F^{-1}XF=e^{2\pi i(\Delta_{\gamma^{-1}})_1}X^dY^{-b},\cr
&F^{-1}YF=e^{2\pi i(\Delta_{\gamma^{-1}})_2}X^{-c}Y^a.\cr}\
\ref{\xyfcomm}
$$

\thm\matrixeltsthm{
\item{(i)} {Under the isomorphism $\kappa$, the action of the
evolution operator F is given by
$$
\kappa F\kappa^{-1}\phi(\theta,z)=F\phi(\gamma\theta+\Delta_\gamma,z),
\ref{\fkappaaction}
$$
with the understanding that, on the right hand side of this equation,
$F$ acts on the $z$ variable.}
\item{(ii)} {The matrix elements of the operator F,
$$
(\phi_m^{({\tilde\theta})},\;F\phi_n^{(\theta)})=
\int_{D}\ol{\phi^{({\tilde\theta})}_m(z)}\;F\phi^{(\theta)}_n(z)\;
d\mu_\hbar(z),\ref{\matrixels}
$$
where
$$
\tilde\theta=\gamma^{-1}\theta+\Delta_{\gamma^{-1}},
$$
are explicitly given by
$$
\eqalign{
(\phi_m^{({\tilde\theta})},\;F\phi_n^{(\theta)})
&=(Nb)^{-1/2}\exp({{i\nu}\over 2})\;\exp{{2\pi i}\over N}
(m\tilde\theta_2-n\theta_2)\cr
&\times\sum_{0\leq r\leq|b|-1}\exp(-2\pi ir\theta_2)\;
\exp{{i\pi}\over{Nb}}\Phi_\gamma(m+\tilde\theta_1,\;n+Nr+\theta_1),\cr
}\ref{\gaussum}
$$
where $\exp(i\nu)=-i\alpha/|\alpha|$, and where
$$
\Phi_\gamma(x,\;y)=ax^2-2xy+dy^2.\ref{\hannber}
$$}}

\proof Part $(i)$ of the theorem is a straightforward
consequence of the previous lemma, and we leave the details to
the reader.

To prove part (ii), we note first the following fact.
If $A,B,C,D,E\in\bC $ are such that ${\rm Re}A>0$, and
$({\rm Re}A)^2>({\rm Re}B+{\rm Re}C)^2+({\rm Im}B-{\rm Im}C)^2$,
then
$$
\eqalign{
&\int_\bC \exp\big(-A|w|^2  + Bw^2 +C\ol w ^2 +Dw+E\ol w\big) dw\,
d{\ol w} \cr
&={{2\pi}\over{\sqrt{A^2-4BC}}}\;\exp{{(D+E)^2(A^2-4BC)-(A(D-E)+
2(EB-CD))^2}\over{4(A-B-C)(A^2-4BC)}}.\cr
}\ref{\gaussianint}
$$
Let $e^{(\theta)}_m(z)$ be the following function:
$$
e^{(\theta)}_m(z)=C_m(\theta)e^{-N\pi z^2+2\sqrt{2}\pi(\theta_1+m)z}.
\ref{\efunction}
$$
Note that
$$
Xe^{(\theta)}_m(z)=e^{2\pi i\theta_1}e^{(\theta)}_m(z),\ref{\xeidentity}
$$
and
$$
Y^ke^{(\theta)}_m(z)=e^{2\pi ik\theta_2}e^{(\theta)}_m(z)
e^{-N\pi k^2+2\sqrt{2}\pi Nkz-2\pi(\theta_1+i\theta_2+m)k}.
\ref{\yeidentity}
$$
As a consequence of \yeidentity,
$$
\phi^{(\theta)}_m=\sum_k\big(e^{-2\pi i\theta_2}Y\big)^ke^{(\theta)}
_m,\ref{\usefulidentity}
$$
an identity which we will find useful later. As a consequence of
\gaussianint, for any $r\in\bZ$,
$$
\eqalign{
\int_\bC\ol{e^{(\tilde\theta)}_m(z)}\;&FY^re^{(\theta)}_n(z)\;
d\mu_\hbar(z)=(Nb)^{-1/2}\exp({{i\nu}\over 2})\;\exp{{2\pi i}\over N}
(m\tilde\theta_1-n\theta_1)\cr
&\times\exp{{i\pi}\over{Nb}}\big(a(m+\tilde\theta_1)^2-2(m+\tilde\theta_1)
(n+Nr+\theta_1)+d(n+Nr+\theta_1)^2\big).\cr}\ref{\importantint}
$$
Consider now the sum:
$$
\eqalign{
\sum_{0\leq r\leq|b|-1}&e^{-2\pi ir\theta_2}
\int_\bC\ol{e^{(\tilde\theta)}_m(z)}\;FY^re^{(\theta)}_n(z)\;
d\mu_\hbar(z)=\cr
&\sum_{0\leq r\leq|b|-1}e^{-2\pi ir\theta_2}
\sum_{k,l}\int_D\ol{X^kY^le^{(\tilde\theta)}_m(z)}\;
X^kY^lFY^re^{(\theta)}_n(z)\;d\mu_\hbar(z).\cr}
$$
Using the commutation relations \xyfcomm\ as well as \xeidentity\
and \usefulidentity, we can rewrite it as
$$
\eqalign{
\sum_{0\leq r\leq|b|-1}&e^{-2\pi ir \theta_2}
\sum_{k,l}e^{2\pi i(-k\tilde\theta_1
+(dk-cl)\theta_1+k(\Delta_{\gamma^{-1}})_1+l(\Delta_{\gamma^{-1}})_2)}\cr
&\times\int_D\ol{Y^le^{(\tilde\theta)}_m(z)}\;
FY^{-bk+al+r}e^{(\theta)}_n(z)\;d\mu_\hbar(z)\cr
&=\sum_{0\leq r\leq|b|-1}\sum_{k,l}\int_D\ol
{(e^{-2\pi i\tilde\theta_2}Y)^le^{(\tilde\theta)}_m(z)}\;
F(e^{-2\pi i\theta_2}Y)^{-bk+al+r}e^{(\theta)}_n(z)\;d\mu_\hbar(z)\cr
&=\sum_{k,l}\int_D\ol
{(e^{-2\pi i\tilde\theta_2}Y)^le^{(\tilde\theta)}_m(z)}\;
F(e^{-2\pi i\theta_2}Y)^ke^{(\theta)}_n(z)\;d\mu_\hbar(z)\cr
&=\int_D\ol {\phi^{(\tilde\theta)}_m(z)}\;
F\phi^{(\theta)}_n(z)\;d\mu_\hbar(z),\cr}
$$
which, together with \importantint\ proves the theorem. $\square$

The sum in \gaussum\ is a generalized Gau\ss\ sum which, for the case
of $\gamma$ of the form \checkerboard\ and $\theta=0$, reduces to the
Gau\ss\ sum studied in [HB].

\section\kronstructsec{The structure of the quantized Kronecker map}

Using the isomorphism we introduced in the previous section, we
construct here the quantized Kronecker dynamics on $\dint_{\bT^2}
\cH_\hbar(\theta)\; d\theta$. In analogy with \matrixeltsthm,
we have the following result.

\thm\kronthm{\item{(i)} {Under the isomorphism $\kappa$, the action of the
evolution operator $K$ is given by
$$
\kappa K\kappa^{-1}\phi(\theta,z)=U(-\omega)\phi(\theta+N\omega,z),
\ref{\kkappaaction}
$$
where $U(-\omega)$ acts on the $z$ variable.}
\item{(ii)} {The matrix elements of the evolution operator $K$,
$$
(\phi_m^{({\tilde\theta})},\;K\phi_n^{(\theta)})=
\int_{D}\ol{\phi^{({\tilde\theta})}_m(z)}\;K\phi^{(\theta)}_n(z)\;
d\mu_\hbar(z),
$$
where
$$
\tilde\theta=\theta-N\omega,
$$
are explicitly given by
$$
(\phi_m^{({\tilde\theta})},\;K\phi_n^{(\theta)})=
\exp\big(2\pi i\omega_2(\theta_1-N\omega_1/2)\big)\;\delta_{mn}.
\ref{\mateltskron}
$$}}

\proof The proof of $(i)$ follows by a simple calculation involving
Fourier series. To prove $(ii)$, we verify by an explicit computation
that
$$
K\phi_m^{(\theta)}=e^{-iN\pi\omega_1\omega_2+2\pi i\omega_2
\theta_1}\phi_m^{(\theta-N\omega)}.\quad\square\num
$$

\medskip\noindent
{\bf Acknowledgments.} The authors would like to thank Eric Heller and
Chris King for insightful comments and criticisms.

\vfill\eject

\centerline{\bf References}
\baselineskip=12pt
\frenchspacing

\bigskip

\item{[AA]} Arnold, V. I. and Avez, A.: {\it Ergodic Problems in Classical
Mechanics}, Benjamin (1968)

\item{[BNS]} Benatti, F., Narnhofer, H., and Sewell, G. L.:
A non-commutative version of the Arnold cat map, {\it Lett.
Math. Phys.}, {\bf 21}, 157--172 (1991)

\item{[BB]} Bouzouina, A., and De Bievre, S.: Equipartition
of the eigenfunctions of quantized ergodic maps on the torus,
preprint (1995)

\item{[C]} Colin de Verdiere, Y.: Ergodicit\'e et functions propres du
Laplacien, {\it Comm. Math. Phys.}, {\bf 102}, 497--502 (1985)

\item{[CFS]} Cornfeld, I. P., Fomin, S. V., and Sinai, Ya.:
{\it Ergodic Theory}, Springer Verlag (1982)

\item{[D]} Degli Esposti, M.: Quantization of the orientation
preserving automorphisms of the torus, {\it Ann. Inst. Poincar\'e},
{\bf 58}, 323--341 (1993)

\item{[DGI]} Degli Esposti, M., Graffi, S., and Isola, S.:
Classical limit of the quantized hyperbolic toral automorphisms,
{\it Comm. Math. Phys.}, {\bf 167}, 471--507 (1995)

\item{[FK]} Farkas, H. M., and Kra, I.: {\it Riemann Surfaces},
Springer Verlag (1980)

\item{[HB]} Hannay, J. H., and Berry, M.: Quantization of linear maps
on a torus - Fresnel diffraction by periodic grating, {\it Physica},
{\bf 1D}, 267--290 (1980)

\item{[KL]} Klimek, S., and Le\'sniewski, A.: Quantized chaotic
dynamics and non-commutative KS entropy, {\it Ann. Phys.}, to
appear

\item{[LV]} Leboeuf, P., and Voros, A.: Quantum nodal points
as fingerprints of classical chaos, in {\it Quantum Chaos}, ed. by
B. V. Chirikov and G. Casati, Cambridge University Press (1994)

\item{[M]} Mumford, D.: {\it Tata Lectures on Theta}, Vol. I,
Birkh\"{a}user (1983)

\item{[S]} Schnirelman, A.: Ergodic properties of the
eigenfunctions, {\it Usp. Math. Nauk}, {\bf 29}, 181--182, (1974)

\item{[Z1]} Zeldich, S.: Quantum ergodicity of $\bC^*$ dynamical
systems, preprint (1994)

\item{[Z2]} Zeldich, S.: Quantum ergodicity of quantized contact
transformations and ergodic symplectic toral automorphisms,
preprint (1994)
\end